# Impact of atmospheric pressure variations on methane ebullition and lake turbidity during ice-cover

Kai Zhao[1], Edmund W. Tedford[1], Marjan Zare[2], and Gregory A. Lawrence[1]

[1]Department of Civil Engineering, University of British Columbia

[2]Department of Mathematics, University of British Columbia

Corresponding author: Gregory Lawrence (lawrence@civil.ubc.ca)

**Abstract**

Methane ebullition (bubbling) from lake sediments is an important methane flux into the atmosphere. Previous studies have focused on the open-water season, showing that temperature variations, pressure fluctuations and wind-induced currents can affect ebullition. However, ebullition surveys during the ice-cover are rare despite the prevalence of seasonally ice-covered lakes, and the factors controlling ebullition are poorly understood. Here, we present a month-long, high frequency record of acoustic ebullition data from an ice-covered lake. The record shows that ebullition occurs almost exclusively when atmospheric pressure drops below a threshold that is approximately equal to the long-term average pressure. The intensity of ebullition is proportional to the amount by which the pressure drops below this threshold. In addition, field measurements of turbidity, in conjunction with laboratory experiments, provide evidence that ebullition is responsible for previously unexplained elevated levels of turbidity during ice-cover.

# 1 Introduction

Lakes are an important source of atmospheric methane (Wik et al., 2016), and ebullition from lake sediments is often the dominant source of this methane (Bastviken et al., 2004; Sanches et al., 2019). However, there are large uncertainties in measuring and predicting methane ebullition (Engram et al., 2020). Multiple environmental factors affect ebullition, including wind-induced currents (Joyce & Jewell, 2003), short wave radiative flux (Wik et al., 2014), sediment temperature variations (Fechner-Levy & Hemond, 1996), water level fluctuations (Harrison et al., 2017), and atmospheric pressure changes (Scandella et al., 2011; Varadharajan & Hemond, 2011). Most studies of ebullition have been conducted during the open-water season, when these processes often act simultaneously, rendering it difficult to discriminate between them. Ice-cover provides the opportunity to study ebullition under less complicated conditions. However, the paucity of ebullition data during ice cover constitutes a major knowledge gap, despite the prevalence of ice-covered lakes (Denfeld et al., 2018; Verpoorter et al., 2014).

In addition to contributing to atmospheric methane, ebullition from sediment can impact the local aquatic system. When bubbles migrate through sediment, they enhance porewater advection (Santos et al., 2015), mobilize and release contaminants (Fendinger et al, 1992), and cause sediment destabilization (Kavcar, 2008). As they emerge from the sediment, bubbles can entrain sediment particles in their wake, increasing turbidity in the water column (Klein, 2006). While ebullition appears to enhance turbidity during ice cover (Lawrence et al., 2016; Tedford et al., 2019a), continuous and simultaneous measurements of ebullition and turbidity have not previously been made.

We examine a one-month record of ebullition collected during ice cover at Base Mine Lake, in Alberta, Canada. During this period, water level fluctuations, temperature changes, and wind-induced currents were minimal, allowing us to focus on the relationship between atmospheric pressure variations and ebullition. A downward facing echosounder was mounted below the ice to monitor ebullition continuously from a fixed area. This high frequency ebullition data revealed the dominant effect of atmospheric pressure on ebullition. We also investigated the impacts of ebullition on lake turbidity with the aid of laboratory experiments and field measurements.

**2 Methods**

Base Mine Lake (57° 1' N, 111° 37' W in Alberta, Canada) was formed by filling a mined-out oil sands pit with tailings and capped with water (Dompierre & Barbour, 2016). As part of a remediation strategy, the deposition of tailings was completed at the end of 2012. The tailings were 25 – 35% solids by weight, with similar mean particle size and clay fraction as the fine-grained muds found in lakes and estuaries (Dompierre & Barbour 2016). These tailings are commonly referred to either as fluid fine tailings, or simply mud. The lake exhibits seasonal thermal stratification similar to that of natural dimictic lakes, and ice-cover is typically from November to April (inclusive) also similar to natural lakes in the region (Tedford et al. 2018).

At the end of 2012, the 7.8 km$^2$ lake was 7 m deep, overlying a 45 m mud layer. The initial depth of the water-cap was designed to isolate the mud from re-suspension due to the direct action of wind-wave induced currents (Lawrence et al., 1991). The mud layer has been settling gradually over time (Dompierre & Barbour, 2016). By January 2018, the average depth of the water-cap was about 10 m (Figure 1a). Degeneration of residual hydrocarbon inside the mud layer produces methane (Francis, 2020); subsequently, bubbles rise through the mud layer and then the water-cap (Figure 1b). These bubbles are typically 0.1 - 1 cm diameter when they reach the water surface.

In the present study we use meteorological data, echo soundings, bottom-water temperature, and turbidity measurements from Day 40 (9 February) to Day 67 (8 March) of 2018 (Zhao et al., 2021). Meteorological data, including atmospheric pressure, have been collected at the central platform (P1) since October 2012 (Figure 1a), and at the Fort MacMurray Airport (369 m ASL), since December 1999 (Environment and Climate Change Canada, Fort McMurray CS). The variations in atmospheric pressure at these two stations were nearly identical (Figure 2a). However, no pressure data were obtained from P1 between Day 40 and Day 47, therefore, we use the airport data for our analysis.

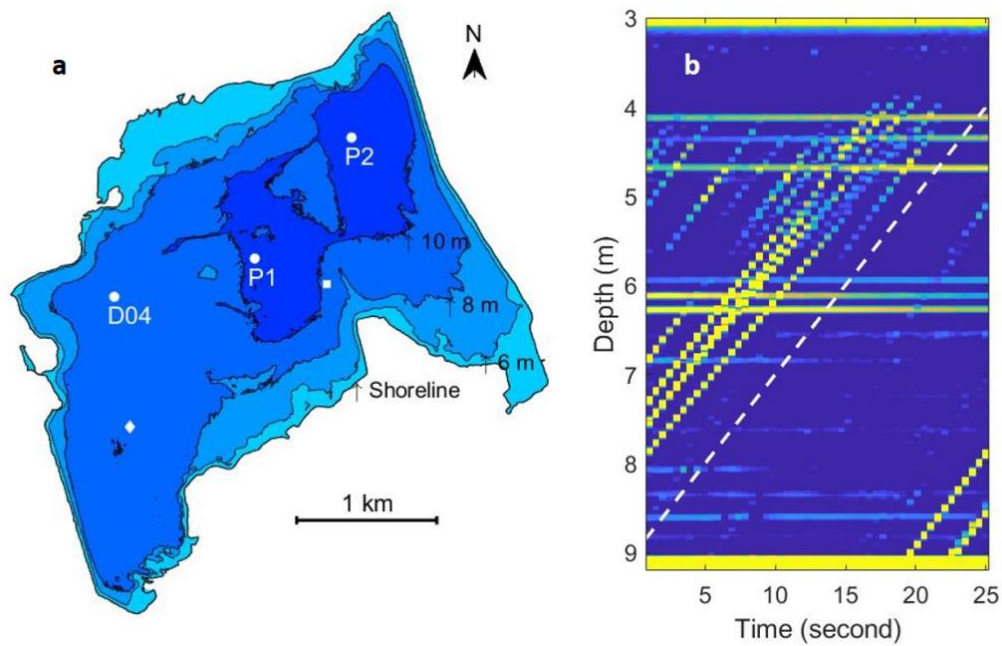

**Figure 1**. Bathymetric map of Base Mine Lake and a sample echogram. (a) shows the lake shoreline and the 6 m, 8 m and 10m depth contours. The white diamond and square mark the locations where images of the lakebed were taken (Figure 4). Meteorological measurements were taken at P1, turbidity measurements and bottom-water temperature at P2, and echo-

soundings at D04. (b) a sample echogram from 25 second acoustic bursts at 09:30, 27 February 2018 (Mountain Standard Time). The echosounder sampled the water column between a depth of 3 m and the lakebed at 9 m. Diagonal lines represent rising bubbles with a speed of roughly 22 cm/sec (the white dashed line indicates a rise velocity of 20 cm/sec). The color of the diagonal lines changes from yellow (stronger backscatter strength) to blue (weaker backscatter strength). The beam width of the echosounder increases with depth, and consequently most of the bubbles leave the beam when they rise above about 4 m depth. Horizontal lines in the water column are the result of floating material. The intense echo (bright yellow) at a depth of 9 m is from the surface of the lakebed.

A downward facing, single-beam echosounder (Echologger EA400) was deployed on 8 February 2018 at D04 (Figure 1a) and recovered after ice-off in May. The instrument was suspended 3 m below the ice and 6 m above the lakebed. Bursts of 50 pings over 25 seconds (2 Hz) were logged once every hour. The instrument was powered by on-board batteries and recorded full profiles of echo intensity internally. With a 5° beam width, the echosounder monitored an area of 0.5 m radius when deployed 6 m above the bottom. A sample echogram is shown in Figure 1b. The stability of the ice and lack of strong currents resulted in particularly clear traces of the rising bubbles (diagonal lines in Figure 1b) as well as a stationary bed (horizontal line at 9 m in Figure 1b). The diagonal lines indicate rising bubbles with a speed of roughly 22 cm/sec. This rise velocity is consistent with the 0.1 – 1 cm diameter bubbles observed at the surface (Clift et al., 2005). To estimate ebullition intensity (Figure 2a) the floating reflectors (the near horizontal lines in the Figure 1b) are filtered out and then the backscatter intensity (decibels) between a depth of 5.2 m and 8.5 m is averaged over the entire 25 second burst. Although the single beam and internal logging of the Echologger EA400 allows for long autonomous deployments under the ice, it does not record bubble locations, or size, inside the beam. Therefore, it is unable to convert ebullition intensity into volumetric methane flux as in the case of dual beam echosounders (Ostrovsky et al., 2008).

Turbidity was measured at 30-minute intervals using a RBRDuo logger with a Seapoint turbidity sensor attached to a mooring chain at P2 (Figure 1a). The total water depth at this location was approximately 11 m. The turbidity in NTU units was approximately twice the total suspended solids concentration in mg/L (Tedford et al., 2019a). On 3 October 2019, images of the lakebed were taken using a drop camera, Subsea Video Systems S-513, to facilitate our understanding of ebullition and turbidity.

## 3 Results

Variations in atmospheric pressure, ebullition intensity and turbidity under ice from 9 February (Day 40) until 8 March 2018 (Day 67) are presented in Figure 2. Ebullition intensity varies dramatically from hour to hour. The most intense ebullition events occurred during the passage of two low-pressure systems. The first system persisted from Day 43 to Day 45, and the second from Day 54 to Day 60 (Figure 2a). During the first event, ebullition started to increase when the atmospheric pressure dropped below its long-term average, increased further as the pressure continued to drop, and then decreased as the pressure increased. Ebullition essentially ceased when the atmospheric pressure rose above its long-term average again. During the second event the pressure cycled down and up three times and ebullition peaked during each pressure trough. During periods of above average pressure, ebullition either ceased altogether, or occurred in occasional bursts with no clear connection to atmospheric pressure.

Increased turbidity (decreased water clarity) at depth is typically associated with low-pressure events (Figure 2c). During the first event (Day 43 – 45) a peak in turbidity at 8.5m coincided with a pressure trough and high ebullition. During the second event (Day 54 – 60) there were three troughs in the atmospheric pressure and three corresponding peaks in ebullition intensity. Turbidity at 8.5m peaked during the second and third pressure troughs, but did not peak during the first pressure trough on Day 54 even though ebullition peaked. Whereas, turbidity at 2.5m was not affected by ebullition or pressure. Note, the bottom-water temperature has no apparent correlation with ebullition (Figure 2d).

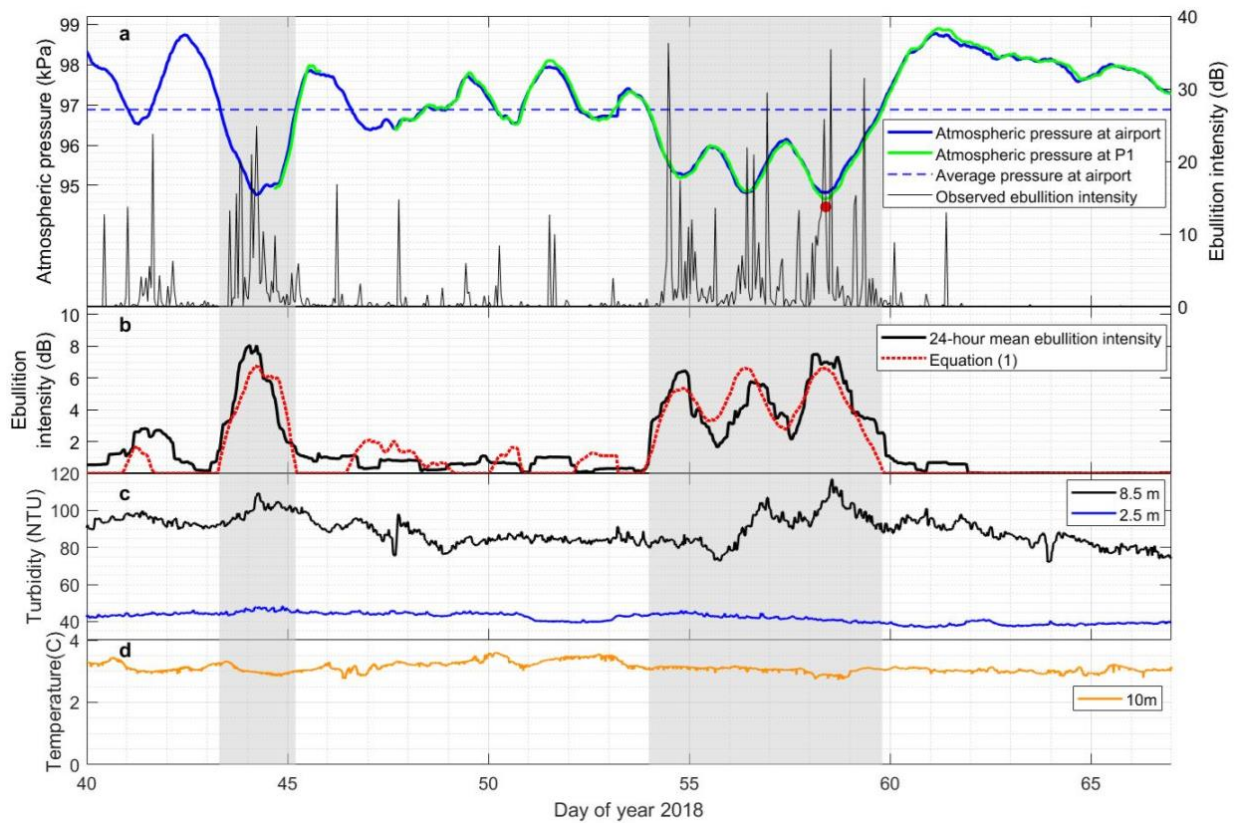

**Figure 2**. Atmospheric pressure, ebullition intensity and turbidity from 9 February to 8 March 2018. (a) The blue line is atmospheric pressure at Fort McMurray airport (47 km south-east of the lake). The green line is atmospheric pressure at P1 in the center of the lake. A calibration

constant of 4.5 kPa has been subtracted from the pressure record at P1. The horizontal dashed blue line is the average pressure of previous three years at the airport. The black line is the observed ebullition intensity obtained from each filtered 25 second echogram. The red dot represents the ebullition intensity estimated from the echogram in Figure 1b. (b) The black solid line is calculated by applying a 24-hour moving average to the observed ebullition intensity. The red dashed line is the ebullition intensity obtained using equation (1). (c) time series of turbidity at 2.5 m (blue) and 8.5 m (black). These depths are relative to the free surface, not the bottom of the ice. (d) time series of water temperature at 10 m depth. The lake at this location is approximately 11 m deep.

We conducted preliminary laboratory experiments to investigate the link between ebullition and turbidity. Following the procedures outlined in Zare and Frigaard (2018), we prepared a layer of Carbopol, a transparent gel-like material, as a surrogate for the mud. We capped the Carbopol with a layer of water to mimic Base Mine Lake (Figure 3). We injected air through a needle, 0.6 mm in diameter, into the base of the Carbopol and observed the rise of the resulting bubbles. As the bubbles rose through the water-Carbopol interface some of the Carbopol, as well as fine particles placed on the interface, were entrained into the wake of the bubbles (Figure 3a). Some of the entrained Carbopol and fine particles detrain before the bubbles rise to the water surface (Figure 3b). We hypothesize that the same process occurs in Base Mine Lake. Turbidity at depth was affected by ebullition (8.5 m in Figure 2c), whereas higher in the water column it was not affected (2.5 m in Figure 2c). After the passage of many bubbles a conduit formed within the Carbopol (Figure 3c). Some of the overlying water (blue) entered the conduit after the release of each bubble.

Photographs of the lakebed (Figure 4) show conduits through which methane bubbles emerge, similar to those observed in the laboratory (Figure 3), and in natural lakes where they cause crater-like depressions in the lakebed "pockmarks" (e.g. Bussmann et al., 2011). These pockmarks are of different sizes and are distributed unevenly over the lakebed. Pockmark 1 (Figure 4a) has an outer diameter of about 1 cm and a smaller inside conduit of approximately 2 mm diameter. Similar conduit sizes have been observed in the laboratory (Kavcar, 2008) and the ocean (Torres et al., 2002). The pockmarks have different depths, diameters and varying spatial distribution. For example, pockmark 8 (Figure 4d) is much larger than pockmark 1, and presumably more gas has passed through it, resulting in greater erosion. On the other hand, pockmark 3 (Figure 4b) is limited in depth and diameter, indicating that it is a younger pockmark. The variety of pockmarks shows that ebullition can gradually modify the morphology of the lakebed. The uneven distribution of pockmarks creates challenges for the monitoring of ebullition. The images in Figure 4 were selected because they have pockmarks, with Figure 4c representing the densest pockmark distribution observed. However, the majority of the images of the lakebed had no pockmarks. Recall that the echosounder only monitors approximately 0.2 $m^2$; if the echosounder were deployed above an area without active pockmarks, ebullition would not likely have been recorded.

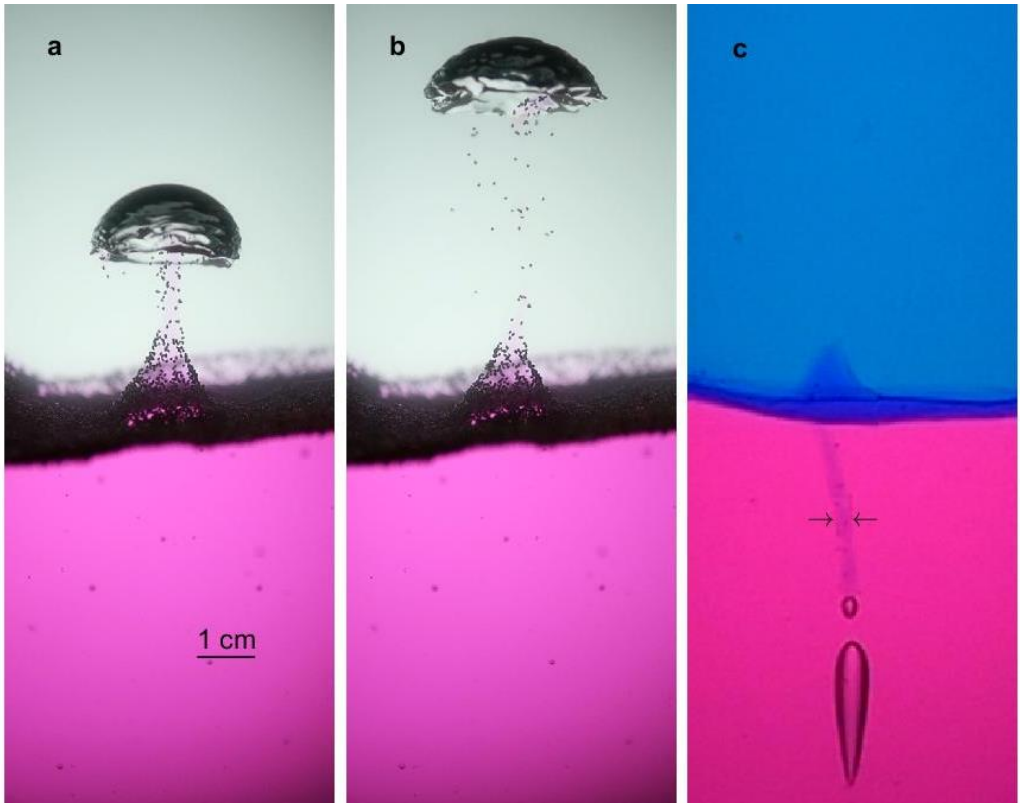

**Figure 3**. Air bubbles rising through a layer of Carbopol into a layer of water. (a) The first in a series of bubbles shown rising through the water after having passed through the Carbopol; its path through the Carbopol is not visible. A cone of Carbopol, as well as fine particles placed on the interface, has been drawn into the wake of the bubble. (b) The same bubble as shown in (a) but 0.09 seconds later. Some of the entrained particles have detrained from bubble wake. (c) An elongated bubble just about to enter a conduit formed in the Carbopol by the passage of previous bubbles. Blue dye has been added to the upper layer to visualize the conduit. The horizontal arrows indicate the width of the conduit. The scale in all three images is the same.

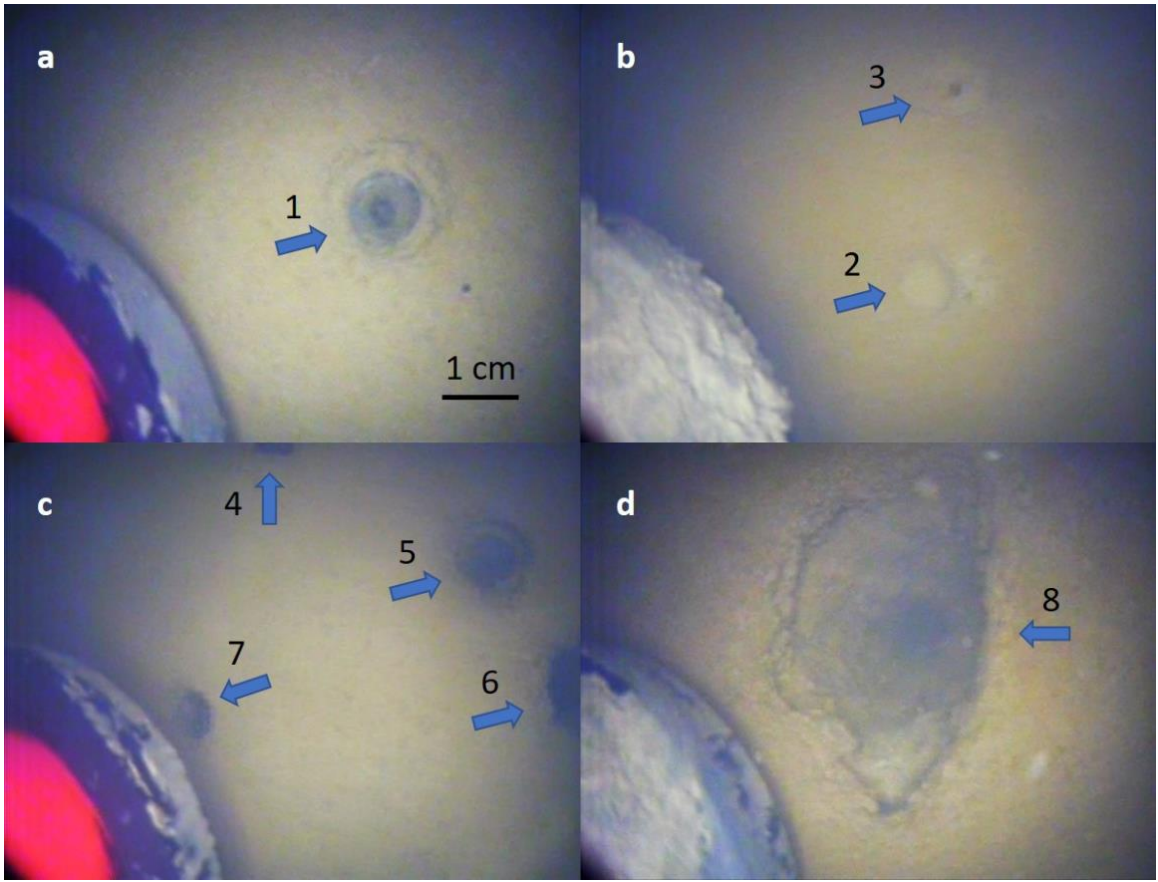

**Figure 4**. Photographs taken from a camera suspended approximately 5 cm above the water-mud interface on 3 October 2019. (a) An active pockmark (#1) with a narrow conduit within. (b) Two pockmarks (#2-3) located close to each other. Pockmark 2 is backfilled with sediment, indicating bubbles did not rise through it in recent ebullition events. Pockmark 3 is particularly small, potentially because few bubbles have passed through it. (c) Four pockmarks (#4-7) in close proximity. (d) One large pockmark (#8). The instrument (red and black color) on the bottom-left corner of each image is an RBR Concerto data logger. The rim of the black guard has a diameter of 8.6 cm. In (b) and (d), the instrument is covered by mud. The location of pockmark 1 is marked as white diamond in Figure 1a; whereas, the location of pockmarks 2 - 8 is indicated with a white square in Figure 1a.

## 4 Discussion

While many researchers have observed a relationship between pressure variations and methane ebullition from sediments (e.g. Walter Anthony et al., 2010; Harrison et al., 2017), to our knowledge none have observed as definitive a relationship as we have presented in Figure 2. Our data set is special for two reasons. Firstly, given that the lake is ice-covered, and there are no inflows or outflows, pressure variations in the sediment are almost exclusively due to atmospheric pressure fluctuations. Secondly, by mounting an echo-sounder in the ice we were able to obtain a continuous record of ebullition from a fixed location for an extended period of time.

In our data record, the ebullition intensity is well approximated by:

$$E_P \approx \begin{cases} k * (P_{th} - P), & \text{if } P < P_{th} \\ 0, & \text{otherwise} \end{cases} \quad (1)$$

where $E_P$ is the predicted ebullition intensity, $P$ is atmospheric pressure, $P_{th}$ is a threshold pressure and $k$ is a proportionality constant. As a measure of the goodness of fit between the measured ebullition, $E_M$, and the ebullition predicted by (1), we use $Q = 1 - \frac{\Sigma (E_P - E_M)^2}{\Sigma E_M^2}$. The variation of $Q$ with $P_{th}$ and $k$ is plotted in Figure S1 in the supporting information. The optimal value of $Q = 0.87$, is obtained when $P_{th} = 97.1 \, kPa$ and $k = 2.9$. Applying this set of values, the result from equation (1) is shown in Figure 2b. It matches the magnitude and timing of observed ebullition well. An alternative set of values is obtained using the average atmospheric pressure $P_{th} = P_{ave} = 96.9 \, kPa$, and $k = 4.2$, calculated by equating the time integral of the observed ebullition intensity over the study period with the time integral of the predicted ebullition intensity using equation (1). This latter pair of values gives $Q = 0.79$ as shown in

Figure S1. A comparison between the observed ebullition intensity and predicted ebullition intensity using this pair of values is shown in Figure S2.

In addition to atmospheric pressure fluctuations, many factors have been suggested to affect methane ebullition. Seasonality in temperature can influence ebullition by affecting methane production rate and methane solubility in porewater (e.g. Fechner-Levy & Hemond, 1996; DelSontro et al., 2016). The temperature of the surface of the mud in Base Mine Lake can vary several degrees on a seasonal basis (Dompierre & Barbour, 2016), in response to the seasonal variation in the temperature of the water column (Tedford et al., 2019a). During the study period, the bottom-water temperature remains stable (Figure 2b) and shows no correlation with ebullition. The temperature inside the mud layer varies even less. Dompierre & Barbour (2016) have shown that temperature variations inside the mud decreases exponentially with depth and the variations are negligible 4 m below the lakebed. Falling water levels, due to reservoir withdrawals, can also trigger ebullition (Harrison et al., 2017). However, due to the absence of inflow and outflow during ice cover, the water level in Base Mine Lake is stable. Finally, Joyce and Jewell (2003) observed that shear stresses caused by bottom currents, driven by surface winds, could trigger the release of bubbles. However, under ice, wind driven bottom currents are negligible.

As in natural sediment, ebullition in Base Mine Lake is subject to complex processes. Nevertheless, based on our observations during ice cover and the effectiveness of equation (1), we conclude that most of the ebullition in Base Mine Lake can be explained using a relatively small number of principles and assumptions, namely: there is a range of depths within the mud of Base Mine Lake in which the methane concentration in the pore water is at, or near, saturation (Francis, 2020); methane bubbles are present in this zone, with sizes ranging from the size at nucleation to a critical size at which ebullition occurs (Algar et al., 2011); and methane released through ebullition is replenished by methanogenesis, but at time scales much longer than the duration of pressure events,.

During ice-cover inflows into, and outflows from, the lake are negligible, and the hydrostatic pressure exerted on the mud by the combination of water and ice is constant. Therefore, any change in atmospheric pressure changes the pressure exerted on the bubbles within the mud. When the pressure experienced by the bubbles drops, the corresponding reduction in methane solubility (Henry's Law) results in gas moving from the pore water into the bubbles, causing them to grow. Decreasing pressure also causes bubbles to grow (ideal gas law). The idealization represented by equation (1) implies that when the atmospheric pressure drops below the pressure threshold (e.g. Figure 2, day 43), the largest bubbles reach critical size and ebullition starts. Subsequently, when atmospheric pressure rises above the pressure threshold (e.g. day 45), the corresponding increase in methane solubility results in methane moving from the bubbles into the pore water. This, together with the reduced volume due to the increased pressure, causes ebullition to stop.

In applying equation (1) we have assumed a constant pressure threshold during our study period, and obtained results that are largely consistent with the observations presented in Figure 2. However, for the pressure threshold to remain constant, the rate of ebullition would need to be in equilibrium with the rate of methane production. If the rate of ebullition exceeds the rate of production the store of available methane will decrease, effectively reducing the pressure threshold; whereas, if production exceeds ebullition the pressure threshold will rise. The relative success of our assumption of a constant threshold in equation (1) implies that the variation of the pressure threshold during our study period was not significant.

While the sediments in Base Mine Lake are the product of a mining operation, they have similar mean particle size and clay fraction as the fine-grained muds found in natural lakes and estuaries (Dompierre & Barbour 2016). The clear linkage between pressure and ebullition that we have observed has also been observed in natural lakes. For example, Matton & Likens (1990) observed rapid ebullition during low atmospheric pressure events in Mirror Lake, New Hampshire. In northern ice-covered lakes, Walter Anthony et al. (2010) also observed that ebullition responded to atmospheric pressure fluctuations. Scandella et al. (2011) showed a strong relationship between pressure variations and ebullition from the "gassy" sediments in Upper Mystic Lake, Massachusetts.

There are occasions (e.g. during days 40, 46, 51 and 61 in Figure 2) when sporadic ebullition occurs even though atmospheric pressure is considerably higher than the pressure threshold. This may be a result of intermittent coalescence of bubbles inside the mud-layer causing them to reach critical size. Movements within the mud layer due to its gradual settling (Dompierre & Barbour, 2016) may facilitate coalescence and ebullition.

Ebullition in Base Mine Lake has other important impacts beyond the elevated turbidity shown in Figure 2c. Our laboratory experiments show that rising bubbles push water out of established conduits, and once the bubbles exit, the water flows back in (Figure 3c). The same process probably occurs in Base Mine Lake, enhancing the exchange of heat and dissolved contaminants between the mud layer and water column. This process provides an explanation for the enhanced mixing hypothesized by Dompierre et al. (2017) in modelling exchange between the mud and water column. The rising bubbles increase dissolved methane concentration in the water column, which leads to methane oxidation and contributes to decreasing dissolved oxygen concentration (Risacher et al., 2018). In winter, methane bubbles are frozen into the ice, along with hydrocarbon transported from the mud (Tedford et al., 2019b). The presence of bubbles causes weakening of the ice, that is apparent during augering, and may result in earlier melting. In summer, hydrocarbon transported by the bubbles reaches the lake surface, influencing heat exchange with the atmosphere (Chang, 2020; Clark et al., 2020) and the properties of wind-generated surface waves (Hurley et al., 2020).

## 5 Conclusions

Although there are extensive observations of ebullition during the open water season, under ice observations are limited. We present high-frequency ebullition data from Base Mine Lake under ice, that reveals that ebullition primarily occurs when atmospheric pressure is below a threshold. This pressure threshold is approximately equal to the average atmospheric pressure. The timing and magnitude of major ebullition events is well reproduced by setting the ebullition intensity to be proportional to the pressure deficit below the pressure threshold. Laboratory experiments and field observations also show that these episodic ebullition events elevate turbidity at depth and can enhance exchange of contaminants between the mud layer and the water column.

Even though our results are from a single lake during ice-cover, the importance of atmospheric pressure variations is not limited to this lake, or the ice-cover season. The response of ebullition to pressure variations has been observed in natural lakes, both during ice-cover (Walter Anthony et al., 2010) and during the open-water season (Matton & Likens, 1990; Scandella et al.,2011). Future studies are needed to refine and extend our results to other lakes and the open-water season. Our data suggests that future surveys need to sample ebullition frequently enough (at least hourly) to capture the rapid response of ebullition to pressure variations.

**Acknowledgement**

The authors are grateful for the support of a Collaborative Research and Development Grant from the Natural Sciences and Engineering Research Council of Canada and Syncrude Canada, Ltd. The authors thank the Syncrude R&D BML Research and Monitoring team and Sarah Chang at UBC for the help in collecting field data, Ian Frigaard at UBC for advice in designing the laboratory experiment, and Roger Pieters at UBC for the valuable input throughout.